\documentclass[3p]{elsarticle}
\usepackage{ecrc}
\volume{}

\firstpage{1}

\usepackage{lineno}
\usepackage{pdflscape}
\usepackage{geometry}
\usepackage{float}
\usepackage{xcolor}
\usepackage{listings}
\usepackage{courier}
\usepackage{multirow}
\usepackage{afterpage}
\usepackage{svg}

\firstpage{1}

\runauth{Thai et al.}

\jid{FGCS}

\usepackage{amssymb}

\usepackage[figuresright]{rotating}
\usepackage[hyphens]{url}
\usepackage{float}
\usepackage{pdfcomment}

\begin{document}

\begin{frontmatter}

\dochead{Accepted to Future Generation Computer Systems, 23 November 2017}

\title{A Survey and Taxonomy of Resource Optimisation for Executing Bag-of-Task Applications on Public Clouds\footnote{This paper can be cited as follows: {\sffamily Long Thai, Blesson Varghese and Adam Barker, ``A Survey and Taxonomy of Resource Optimisation for Executing Bag-of-Task Applications on Public Clouds,'' Future Generation Computer Systems, ISSN: 0167-739X, Elsevier Press, Amsterdam, The Netherlands, 2017 (in press).}}}

\author{Long Thai}
\address{School of Computer Science, University of St Andrews, Fife, UK}
\ead{long.t.thai@gmail.com}

\author{Blesson Varghese}
\address{School of Electronics, Electrical Engineering and Computer Science, Queens University Belfast, Belfast, United Kingdom}
\ead[url]{http://www.blessonv.com}
\ead{varghese@qub.ac.uk}

\author{Adam Barker\corref{mycorrespondingauthor}}
\address{School of Computer Science, University of St Andrews, Fife, UK}
\cortext[mycorrespondingauthor]{Corresponding author}
\ead[url]{http://www.adambarker.org/}
\ead{adam.barker@st-andrews.ac.uk}

\begin{abstract}

Cloud computing has been widely adopted due to the flexibility in resource provisioning and on-demand pricing models. Entire clusters of Virtual Machines (VMs) can be dynamically provisioned to meet the computational demands of users. However, from a user's perspective, it is still challenging to utilise cloud resources efficiently. This is because an overwhelmingly wide variety of resource types with different prices and significant performance variations are available.

This paper presents a survey and taxonomy of existing research in optimising the execution of Bag-of-Task applications on cloud resources. A BoT application consists of multiple independent tasks, each of which can be executed by a VM in any order; these applications are widely used by both the scientific communities and commercial organisations. The objectives of this survey are as follows: (i) to provide the reader with a concise understanding of existing research on optimising the execution of BoT applications on the cloud, (ii) to define a taxonomy that categorises current frameworks to compare and contrast them, and (iii) to present current trends and future research directions in the area.
\end{abstract}


\end{frontmatter}

\section{Introduction}
\label{sec:introduction}

\emph{Cloud computing} has become a sizeable industry and allows users, including industry and academic organisations to rent resources. According to a Business Insider report, the revenues of Amazon Web Services (AWS) and Microsoft Azure have exceeded \$6 billion a year \cite{business_insider}. The adoption rate of cloud computing is high, according to a report by RightScale \cite{rightscale}; nearly 88\% of 930 organisations considered in the report took advantage of cloud computing. 

There are usually two parties involved in cloud computing: \emph{cloud providers} and \emph{cloud users}. Cloud providers build and maintain data centres, such as Amazon or Google. They manage and maintain the physical infrastructure on which the cloud is running and define the types of resources that are available to users and the associated pricing. 

Cloud users require computational resources to run applications. Users range from commercial companies, academic organisations, or private users who deploy and run a few or all of their applications on the cloud. Therefore, cloud users do not need to pay attention to the deployment and maintenance of the physical infrastructure.
However, they are responsible for utilising the resources offered by the providers to build their own \emph{cloud cluster}. We define a cloud cluster as a collection of VMs that is used to achieve the intended goals of a workload deployment. 

Research in cloud computing can be broadly classified on two different points of view, both of which are necessary for developing next-generation cloud computing systems~\cite{Varghese2017}. The first one aims to help cloud providers efficiently build, manage and operate cloud infrastructure. Research in this direction can be categorised as \emph{cloud data centre optimisation}, in which the resources are represented as Physical Machines (PMs) that a cloud provider owns and maintains. The optimisation technique, referred to as \emph{VM placement}, aims to map VMs onto PMs in order to minimise the number of allocated PMs.

The second category is based on \emph{cloud usage optimisation}, which takes the user's point of view into account and deals with optimisation tasks such as: how does a user make a decision about which resources to utilise, or when to scale an application on the cloud. Inputs describing the cloud environment, a user's application(s) and requirements are taken into account to determine a course of action, such as resizing a cloud cluster and distributing workloads among VMs. Since the physical infrastructure is abstracted away from the user, research in this direction normally assumes that resources are unlimited and focuses on minimising the incurred monetary costs associated with renting VMs.

Users run a variety of applications or workloads on the cloud, ranging from simple Web applications, workflows and frameworks which support computationally-intensive applications, such as MapReduce, and Spark. A Bag-of-Task (BoT) application is one class of workload that is commonly used on the cloud and consists of many independent tasks, each of which can be executed by any machine in any order. They can be executed concurrently by many different machines. For instance, a simulation application, e.g. Monte Carlo simulation~\cite{WICS:WICS1314}, is a BoT application in which each execution represents a task. Similarly, a parameter sweep application~\cite{Casanova:2000} is a BoT application in which each task corresponds to one combination of parameters. However, a MapReduce~\cite{Dean:2008} application is not a BoT application since the Reduce phase must wait for the Map phase to complete. The Map and Reduce phases can be considered as two different BoT applications. This survey paper focuses on the cloud usage optimisation for BoT applications.

We have selected to survey BoT on the cloud because they are widely utilised by both scientific and commercial organisations. These applications are large and too complex to be executed on a single machine. They are also the dominant applications that are submitted to and utilise CPU time in grid environments~\cite{Iosup:2011}. Similarly, companies, such as Facebook, report that the jobs running on their own internal data centres are mostly independent tasks~\cite{Bistro}.


Even though there have been multiple surveys regarding research in cloud optimisation, we believe that they do not provide a holistic view of the research in cloud usage optimisation. For instance, the surveys of Fakhfakh et al.~\cite{Fakhfakh:2014} and Wu et al.~\cite{Wu:2015} focus on workflow applications. Surveys in this area are usually based on the point of view of cloud providers~\cite{Faniyi:2015} or treat optimising cloud usage as one of many aspects of cloud data centre management~\cite{Mann:2015}. 

This paper is distinguished from existing surveys by focusing on the methodologies that can be used by cloud users, who do not have a complete view of the underlying infrastructure of the public cloud. In this survey, we set out to review the existing publications regarding optimising the cloud resource from a user's point of view. Furthermore, we focus only on BoT applications. 

The goal of this survey is threefold: (i) to provide a holistic and concise view of the current state-of-the-art in cloud usage optimisation for BoT applications, (ii) to define a taxonomy that categorises current research to compare and contrast existing frameworks, and (iii) to present current trends and employ the taxonomy as a guide for furthering research in the area. 

\subsection{Data Collection}
The research publications used in this survey were collected in September 2017 via Google Scholar. To ensure the quality of the publication, we selected articles from high-impact journals, such as IEEE Transactions on Services Computing and IEEE Transactions on Parallel and Distributed Systems, and top-tier conference venues, such as IEEE Conference of Cloud Computing (CloudCom), IEEE/ACM International Symposium on Cluster, Cloud and Grid Computing (CCGRID), and IEEE International Conference on Cloud Computing (CLOUD).

We set out the following criteria for a publication to be selected for this survey:
\begin{itemize}
    \item The application presented in the publication must be a BoT application; we did not consider publications that presented workflow or user-facing applications.
    \item The execution results presented in the publication must be performed fully or partly on the cloud; we did not consider other resource environments, such as grids.
    \item The monetary cost incurred in executing an application must be considered in the publication, which is a unique characteristic of the cloud environment.
    \item The assumption in the publication must be that the cloud is a black-box environment, e.g., a public cloud in which a user has little to no control over internal operations.
\end{itemize}

The above criteria were set in line with our survey goals - develop a taxonomy of resource optimisation for executing BoT applications on public clouds. At the end of the publication selection phase, there were 31 publications that satisfied the criteria and are used as the basis for this survey.

\subsection{Organisation}

\begin{sidewaysfigure}
      \centering
      \includegraphics[width=\textwidth]{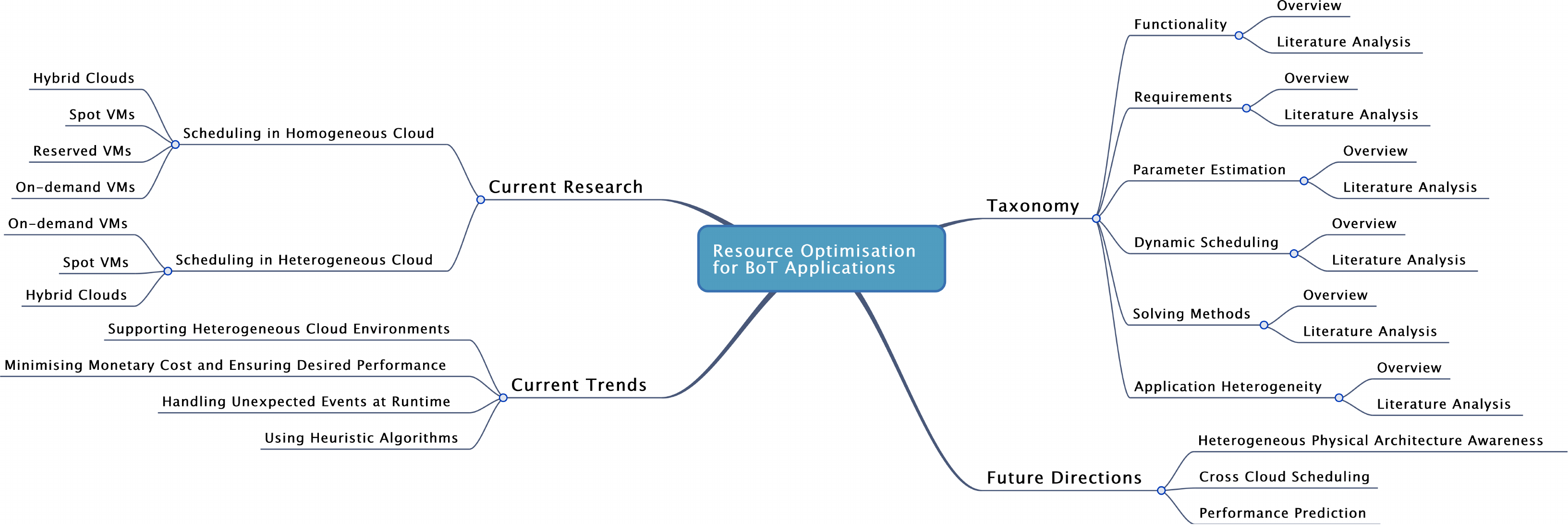}
      \caption{Overview of Resource Optimisation for Executing Bag-of-Task Applications on the Cloud}
      \label{fig:overview}
  \end{sidewaysfigure}

The organisation of this survey is shown in Figure~\ref{fig:overview}. 
We firstly consider the current research of BoT applications in Section~\ref{sec:existing}. As previously indicated we present this from the perspective of the methodologies that a cloud user can adopt rather than the techniques used in the underlying infrastructure or middleware of public clouds that is usually inaccessible to a user. Both scheduling of BoT applications on a \emph{homogeneous} cloud and a \emph{heterogeneous} cloud are considered. We refer to homogeneous clouds as environments that use the same type of VMs in public clouds, and to heterogeneous clouds as environments where different VM types are used. We explore scheduling in the context of hybrid clouds, spot VMs, and on-demand VMs for both homogeneous and heterogeneous clouds and in addition for reserved VMs in homogeneous clouds.

The taxonomy we propose in Section~\ref{sec:taxonomy} is based on six themes, namely functionality, requirements, parameter estimation, dynamic scheduling, solving methods and application heterogeneity. For each of these themes, we first present an overview and then the associated review of the literature. Our survey then uses the above taxonomy for summarising four current trends that are seen in BoT scheduling, Section~\ref{discussion:currenttrends}. We use these to chart out three future directions for optimising cloud usage for executing BoT applications in Section~\ref{discussion:futuredirections}. 

Although the structure presented in Figure \ref{fig:overview} is created to survey research specific to scheduling BoT applications, it may be broadly used for other applications, such as workflows or user-facing applications.

\section{Current Research}
\label{sec:existing}
In this section, we survey research that focuses on developing frameworks for scheduling the execution of BoT jobs on the cloud. This survey is focused on BoT jobs on cloud resources and therefore alternate application types (such as workflows) and resource models (such as the grid or clusters) are not considered. This survey assumes a cloud user's point-of-view, which treats the cloud as a black box and the user may not have control over its internal operations. For example, in a public cloud, a user may be oblivious to the scheduling technique to allocate a VM on a physical machine running in a data center. So all publications that either require the knowledge or affect the internal structure of the cloud are not considered in this survey. For instance, we exclude all publications on energy efficiency since a user cannot influence this on a public cloud. 

This survey considers both \emph{homogeneous} and \emph{heterogeneous} cloud environments. We define a cloud environment to be homogeneous if every VM in a cloud cluster is of the same pre-defined instance type. We define a cloud environment to be heterogeneous if VMs of different instance types are available for an application. 

\subsection{Scheduling in Homogeneous Cloud Environments}

In homogeneous cloud environments, given that only one instance type is used to create VM, both the expected performance and pricing are the same on all available VMs. This simulates an ideal data center and simplifies optimisation.

\subsubsection{Hybrid Clouds}
In a hybrid cloud, both private and public clouds are used. Even though each of them normally has different instance types, we still consider the following publications as using homogeneous cloud since they only use one instance type of the public cloud. In other words, the incurred monetary cost is uniform.

Candeia et al. \cite{Candeia:2010} proposed a framework that schedules an application such that the deadline is met while monetary costs are minimised. The scheduling problem is modelled to simulate different scenarios by determining the number of public cloud VMs to be rented. The scenario that results in the highest profit is selected.

Bicer et al. \cite{Bicer:2012} not only considered the monetary cost of renting cloud VMs, but also the overhead for synchronising the private and public clouds. For example, transferring data between two clusters. A mathematical model for predicting the execution time and the total cost of a hybrid cloud cluster is used. This model calculates the number of VMs to be rented from public cloud providers in order to satisfy the deadline or budget constraints.

Duan and Prodan \cite{Duan:2014:CloudCom} introduced a game theoretic approach to solving a multi-objective scheduling problem which aimed to minimise not only the cost but also the execution time while satisfying bandwidth and storage constraints. The proposed heuristic algorithm first optimised the performance based on given bandwidth constraint. When the range of possible solutions was found, cost optimisation was applied to select the optimal solution. The result was the \emph{task distribution matrix} which defines the distribution of BoT tasks on private and cloud VMs.

Similarly, Hoseinyfarahabady et al. \cite{Hoseinyfarahabady:2013} proposed a heuristic algorithm which aimed to minimised the makespan and cost of executing a BoT application on hybrid cloud. The authors assumed that task execution time was not available prior to the execution. So the execution of BoT tasks was divided into different time intervals and the results from all intervals were combined to estimate the task execution time. 

\subsubsection{Spot VMs}
The performance of a cloud environment is normally improved by using preemptible VMs, also referred to as spot VMs. These VMs are obtained through a bidding process and may be terminated by the provider without any notice. The pricing of spot VMs is normally lower than on-demand VMs. However, the price may fluctuate dynamically over time based on the number of bidders. In this context, research in scheduling focuses on finding an effective bidding strategy for scheduling applications or managing an application in the event of sudden termination, or both.

Yi et al.~\cite{Yi:2012} developed a checkpointing mechanism that saves the progress of application execution at different points in time. This minimises the amount of execution time an application would lose if the VM is suddenly terminated. The framework monitors the bid prices in real-time in order to predict a termination. When such an event is predicted, the current process is saved.

Instead of using fixed bid prices, AMAZING~\cite{Tang:2012} uses Constrained Markov Decision Process to find an optimal bidding strategy. The proposed approach takes deadlines into account and calculates the probability of different bidding options. When a predicted bidding price is too high, the framework saves the current process and waits for the next billing cycle to bid again.

Lu et al.~\cite{Lu:2013} used spot resources for executing BoT jobs. The authors focus on the robustness of the system by using on-demand VMs, which are usually more expensive. However, the impact of termination is minimised since on-demand VMs are used as a backup. Whenever spot instances are terminated, the workload is immediately offloaded onto on-demand VMs.

Menache et al.~\cite{Menache:2014} suggests switching to on-demand resources when there is no spot instance available to ensure the desired performance is always achieved. The decision to use on-demand resources is based on either a user-defined deadline or a policy to allocate a fixed number of on-demand VMs.

\subsubsection{Reserved VMs}
Costs in cloud environments can be reduced by using reserved VMs. This requires upfront payment for the VM but generally the resources are available at lower costs than on-demand VMs. This pricing scheme is useful if a user has a long-term plan regarding the usage of the resource. Overprovisioning, which is when a user reserves more resources that are not fully utilised may be a problem that will need to be tackled when reserving VMs. To mitigate this, cloud environments consisting of both reserved and on-demand VMs are employed. A significant proportion of the workload is assigned to reserved VMs to increase their running time. On-demand instances may be added in order to temporarily handle resource bursts in the workload.

Yao et al.~\cite{Yao:2014} presents an approach for satisfying job deadlines while minimising monetary cost by using both on-demand and reserved VMs. Heuristic algorithms that aim to pack as many jobs as possible into reserved VMs are proposed for increasing utilisation during the lease period. The remaining jobs are assigned to on-demand VMs. This resulted in achieving the desired performance at the lowest cost.

Shen et al.~\cite{Shen:2013} uses reserved VMs to optimise cloud environments to achieve cost savings. Integer Programming is used to model the assignment of tasks on VMs and the cost is minimised by determining the number of reserved and on-demand VMs. This scheduling problem is solved periodically in order to take into account newly submitted workloads.

\subsubsection{On-demand VMs}
Thai et al.~\cite{Thai:2014:CloudCom} proposed a framework for scheduling BoT applications in which tasks are distributed across different geographic locations. This allows for reducing data transfer time by placing application tasks closer to the source of data. A heuristic algorithm is used to this end and then tasks are reassigned between VMs to reduce the number of VMs employed in the environment. Consequently, there are cost savings, but at the same time satisfies a user-defined budget. This research is also extended to make use of idle VMs (these VMs were executing tasks, but complete execution before other tasks complete execution~\cite{Thai:2015:Closer}) to minimise VM under-utilisation and total execution time.

\subsection{Scheduling in Heterogeneous Cloud Environments}
We now consider research that makes use of a wider variety of VM types (or heterogeneous resources) offered by providers.
In this section, we present existing methodologies for scheduling in heterogeneous environments. Compared to homogeneous cloud environments, a heterogeneous environment can be designed to offer more flexibility. This is conducive for applications that have a preference on the hardware specification or configuration of the VM. However, this is more challenging and the framework must take into account the trade-off between cost and performance of different VM types.

\subsubsection{Hybrid Cloud}
Wang et al.~\cite{Wang:2016} proposed a framework that incorporates a heuristic algorithm which greedily assigns tasks to the best performing physical machine in a private cloud cluster. However, if no physical machine is available on the private cloud, the framework provisions VM from a public cloud based on a user-defined budget.

Van Den Bossche et al.~\cite{Bossche:2013} uses priority queues for scheduling BoT execution on the hybrid cloud. Each job is associated with a specific deadline and is added to a queue when it is submitted. A mechanism that periodically scans the queue and estimates if the jobs can meet their deadlines using a private cluster is developed. If this is not possible, a job is moved onto a VM with the cheapest VM type.

Kang et al.~\cite{Kang:2013} proposed a framework that minimises the cost of a hybrid cloud for executing BoT job with a deadline.
Tasks are first considered to be executed on either private machines or existing cloud VMs since they do not incur monetary costs. However, if no existing resources are available, then the framework selects the cheapest VM which can execute the tasks within the deadline.

Duan et al.~\cite{Duan:2014} employs a game theory based scheduling on hybrid clouds. A multi-objective scheduling mechanism in which not only the makespan and monetary cost are minimised, but also the bandwidth and storage limit are not exceeded is proposed.
A \emph{K-player cooperative game} approach in which the players represent the applications that share the same private cluster is used. The algorithm aims to assign both private and public resources to each player so that the makespan and cost are kept to a minimum while the storage and bandwidth constraints are satisfied.

Pelaez et al.~\cite{Pelaez:2016} argue that a scheduling framework should also take into account the variation of task execution during runtime. Therefore different approaches to estimate the task execution time during execution is used to constantly update the scheduling plan. Based on the updated information, tasks are assigned to the cheapest VMs that can execute the tasks within the deadline by taking into account the variation of task execution time.

\subsubsection{Public Clouds with Spot VMs}
Chard et al.~\cite{Chard:2015} employ spot resources to achieve cost savings. An iterative process repeatedly bids for spot VMs to execute jobs. The maximum bid price is always kept lower than the on-demand resource price. However, if there is a job that is waiting for more than a predefined amount of time, it will be executed using the cheapest on-demand VM that can execute a task within the deadline.

\subsubsection{Public Clouds with On-demand VMs}

There is research that focuses on executing a single BoT application. Oprescu et al.~\cite{Oprescu:2010} present BaTS which is a budget-constrained scheduler for executing BoT job on the cloud. The problem is modelled as a Bounded Knapsack Problem and is solved using dynamic programming. The objective is to identify the number of VMs of each type for an application so that the total monetary cost does not exceed the budget constraint while not compromising performance. This research is extended to include the replication of tasks from running VMs onto idle VMs with the intention of decreasing the overall execution time~\cite{Oprescu:2012}. 

Ruiz-Alvarez et al.~\cite{Ruiz-Alvarez:2015} model the problem of minimising the cost of executing BoT jobs on the cloud using Integer Linear Programming. The execution of the application is divided into multiple intervals, each of which might correspond to one billing cycle (for example, one hour). In order to execute all tasks within a deadline, the number of tasks that are required to be executed within each cycle is estimated. Then the model selects the number of VMs so that all tasks are executed within the interval.

HoseinyFarahabady et al.~\cite{HoseinyFarahabady:2014} focus on the trade-off between performance and cost in scheduling BoT applications on the cloud. This trade-off represents a user's preference. For instance, a user might want to achieve high performance while knowing that it would result in a higher monetary cost. For this an algorithm using the Pareto frontier is employed by distributing tasks onto VMs of different types for execution.

Thai et al.~\cite{Thai:2015:Cloud:BoT} proposed a heuristic algorithm for executing a BoT application given either budget or deadline constraints. 
A homogeneous environment is iteratively transformed into a heterogeneous environment by replacing existing VMs of the same type with different VM types. The goal again is to reduce either the total cost (if a deadline constraint is given) or the execution time (if a budget constraint is given) without violating the constraint.

There is also research that considers the execution of multiple BoT applications. 
Mao et al.~\cite{Mao:2010} propose an approach to schedule the execution of multiple BoT jobs on the cloud with both deadline and budget constraints. In this approach, prior knowledge (i.e. the number of tasks of each job that a VM of a certain type could execute within an hour). The scheduling problem is then modelled as an Integer Programming problem and generates a plan with the number of VMs of each type that can meet both deadline and budget constraints. Scheduling is performed periodically at the end of each billing cycle.

Lampe et al.~\cite{Lampe:2012} determined the mapping between BoT jobs and VMs so that all jobs can be executed within their deadlines with a minimum cost. Two different approaches are proposed for solving the problem. The first approach is modelled as a Binary Integer Problem and the second approach is based on heuristic algorithms. The latter approach repeatedly selects the cheapest VM to execute a list of jobs. Based on simulation studies, it is observed that the approaches require a significantly large amount of time to find a solution that can reduce the overall costs.

Gutierrez-Garcia et al.~\cite{Gutierrez-Garcia:2013} presented a policy-based approached for scheduling BoT execution on the cloud. A portfolio of 14 heuristic algorithms, each of which use a different task ordering and resource mapping policy. Experimental results indicate that the effectiveness of the algorithm depends on the characteristics of the workload.

Zou et al.~\cite{Zou:2014} employ a Particle Swarm Optimisation (PSO) technique to execute multiple BoT jobs with a deadline on the cloud while minimising the cost. Additional constraints in terms of the number of CPU cores and the amount of memory each job requires is considered. The traditional PSO technique is compared with a self-adaptive learning PSO (SLPSO) which has greater chances of finding either a better local optimal or even a global optimal.

Thai et al.~\cite{Thai:2015:Cloud:BoT} developed a mechanism for scheduling multiple BoT applications given a budget constraint. Multiple homogeneous cloud environments are merged to create a single heterogeneous environment. The size of a cluster is then reduced by removing VMs from the cluster by reassigning tasks of a VM onto other VMs that have spare resources. A combination of mathematical optimisation and heuristic algorithms are also used~\cite{Thai:2016:DIDC}. A complete optimisation approach is also employed for determining an optimal scheduling plan~\cite{Thai:2016:CloudCom}. 

OptEx~\cite{Sidhanta:2016} is a scheduling framework built on Apache Spark~\cite{spark}. The framework does not require prior knowledge regarding the execution time of a job on a VM. Instead, this knowledge is acquired by profiling the execution of the job to construct a prediction model that estimates a job completion time based on the number of VMs in the environment. This estimation is used by a Non-linear Programming model to calculate the number of VMs of each type that are required to execute a job within a deadline. Workload assignment is performed using a built-in Spark mechanism.

\section{Taxonomy}
\label{sec:taxonomy}

This section identifies common themes, characteristics, requirements, and challenges based on the publications surveyed in the previous section. This taxonomy is shown in Figure~\ref{fig:overview} by illustrating different characteristics and categories of a cloud scheduling framework.

The taxonomy is based on the following six themes. 
\begin{itemize}
\item \textbf{Functionality (refer Section~\ref{tax:func})}: defines the functionalities that are available in scheduling techniques proposed in literature. The three basic functionalities we identify are methods to (i) select the instance types for scheduling a cloud cluster, (ii) scale cloud VMs for each instance type, and (iii) allocate workloads to VMs running in a cloud cluster. 
\item \textbf{Requirements (refer Section~\ref{tax:requirements}):} defines the criteria set out to evaluate the quality of execution - whether it has been successful or not. We consider constraints, which are the criteria that must be satisfied and objectives, which measure the quality of execution on the cloud. 
\item \textbf{Parameter Estimation (refer Section~\ref{tax:estimation}):} defines the variables that need to be estimated during runtime for efficient scheduling. These include monetary and performance factors. 
\item \textbf{Dynamic Scheduling (refer Section~\ref{tax:dynamic}):} defines the estimation of parameters for re-scheduling the application since its initial deployment on the cloud. We consider how dynamic scheduling is triggered and how it is performed in the cloud. 
\item \textbf{Solving Methods (refer Section~\ref{tax:solving}):} defines the techniques used to obtain a scheduling plan. Given that a scheduling plan is the solution to an optimisation problem, we identify that exact algorithms and heuristic algorithms are employed. 
\item \textbf{Application Heterogeneity (refer Section~\ref{tax:hete}):} defines the techniques for scheduling multiple BoT applications each of which has different requirement and performs differently.
\end{itemize}

\subsection{Functionality} \label{tax:func}

\subsubsection{Overview}

The functionality of a scheduling approach defines what it does, i.e. the action(s) that it performs.
There are three basic functionalities that may be combined to create more sophisticated functionalities:

\begin{enumerate}
    \item \textbf{Type selection:} involves determining the combination of instance types that are used in the cloud cluster. Scheduling frameworks must be aware of the difference in not only prices but also performance across all instance types.
    \item \textbf{Resource scaling:} calculates the number of VMs for each instance type. This functionality directly affects the incurred monetary costs (as more VMs are added to the cloud cluster for an application, the more expensive it will be).
    \item \textbf{Workload allocation:} functionality assigns workloads to the VMs running in the cloud cluster. The allocation needs to take into account the performance of a VM, its current state (i.e. knowledge of the workload already on a VM), and an application's requirements.
\end{enumerate}

\subsubsection{Literature Analysis}

Obviously, type selection is not covered by methodologies that only support a homogeneous cloud cluster, for example, a cluster where all resources consist of the same VM type. However, a user must decide in advance which instance type is the most suitable for an application.

The most straightforward method for type selection is to run the application on a few or even all available instance types in order to determine which instance would be the most suitable (this is suggested by AWS~\cite{aws_instance_types}). However, this method can be not only time-consuming but also expensive, due to the number of available instance types and applications. Researchers have proposed approaches to find the most suitable instance type of a user's applications. For instance, Varghese et al.~\cite{Varghese:2014} proposed a framework for instance type selection by matching VM performance, obtained via benchmarking, with an application's characteristics.

Some researchers have chosen to exclude workload allocation.
Since Sidhanta et al.~\cite{Sidhanta:2016} built their OptEx framework on the Spark framework, they relied on the existing built-in mechanisms for workload allocation.
On the other hand, Mao et al. \cite{Mao:2010} assumed prior knowledge regarding the fixed distribution of task allocations to each VM type.
Finally, other researchers have chosen to describe the required performance of a job as the number of CPU hours and degree of parallelism, which eliminates the need for workload allocating~\cite{Menache:2014,Yao:2014}.

The \textbf{resource selection} process, which consists of resource scaling and type selection (if a cloud cluster is heterogeneous) can be performed separately from the workload allocation process.
More specifically, the former will first determine the total amount of resources within the cloud cluster, then the latter will assign the workload to each VM.
This approach usually simplifies the scheduling process.
For instance, many researchers have adopted a mechanism in which the workload is sequentially assigned to the first idle instance~\cite{Candeia:2010,Bicer:2012,Yi:2012,Tang:2012,Oprescu:2010}.
On the other hand, the research of Thai et al. performed the resource selection first and then used different mechanisms for workload allocation~\cite{Thai:2016:DIDC}.

Finally, the majority of the existing work treats resource selection and workload allocation as an interrelated process~\cite{Lu:2013,
Shen:2013,
Thai:2014:CloudCom,
Thai:2015:Closer,
Wang:2016,
Bossche:2013,
Kang:2013,
Duan:2014,
Pelaez:2016,
Chard:2015,
Oprescu:2012,
HoseinyFarahabady:2014,
Thai:2015:Cloud:BoT,
Lampe:2012,
Gutierrez-Garcia:2013,
Zou:2014,
Thai:2016:CloudCom,
Thai:2015:Cloud:BoTs}. In other words, the decision-making process must consider not only a number of required resources but also the allocation of the workload onto those resources. More specifically, the workload can only be assigned to VMs that are created. Similarly, VMs must only be created if there will be workload assigned to them.
We believe that this is the most demanded requirement for running an application on the cloud, due to the incurred monetary costs which do not occur in any other cluster systems like the grid or in-house data centres. As a result, it is necessary to provide users with a framework that helps them keep track of the cost in order to avoid unnecessary spending.
Furthermore, achieving the desired level of performance is really important for running any type of application. For BoT applications, the desired performance is normally represented as a deadline. The violation of deadline constraints can lead to undesired consequences for the user such as financial penalties or a customer's dissatisfaction with the service~\cite{Jockey}.

\subsection{Requirements} \label{tax:requirements}

\subsubsection{Overview}

If the functionality of a scheduling approach defines what it does, its \textbf{requirements} describe why it does what it does, i.e. its goals.
Basically, they are the set of criteria that is used to either determine if the execution of applications on the cloud is successful or evaluate the quality of the execution.
The requirements are set by a user and it is the goal of any cloud usage optimisation methodology to satisfy those requirements.

\subsubsection{Literature Analysis}

Requirements can be divided into two categories.
\textbf{Constraints} are criteria that must be satisfied. Failure to satisfy those constraints is normally unacceptable.
For instance, the research of Thai et al. \cite{Thai:2015:Cloud:BoT} aimed to satisfy either budget or deadline constraints.
Occasionally, it is possible to violate a constraint given there is no other solution. However, the violation of such constraint, i.e. \emph{soft constraint}, should be kept minimal. A constraint is normally represented as a threshold with a specific value such as a deadline or budget constraints \cite{Bicer:2012,Mao:2010}.

On the other hand, \textbf{objectives} are used to measure the quality of an application's execution on the cloud.  For instance, if cost saving is an objective, the cheaper the execution is the better. An objective is normally represented as a goal to minimise, or maximise one or more parameters, e.g. minimise the monetary cost.
Since using cloud resources incurs monetary costs, it is necessary for a user to be aware of the cost that he/she has to pay. As a result, one of the most common requirements for optimising cloud usage is \emph{cost minimisation}~\cite{Candeia:2010,
Yi:2012,
Lu:2013,
Shen:2013,
Wang:2016}.
On the other hand, \emph{performance maximisation} is an objective, which aims to minimise an execution makespan \cite{Duan:2014,Gutierrez-Garcia:2013}.
There are other more specific objectives such as the trade-offs between performance and cost \cite{HoseinyFarahabady:2014}. In this research, a user was asked to provide a numerical value which, represented his/her preferences over cost or performance.

The majority of the existing work considers both constraints and objectives whilst making a scheduling decision. For instance, some researchers have focused on optimising cloud usage by \emph{minimising the monetary cost while satisfying the deadline constraint}~\cite{Tang:2012,
Menache:2014,
Yao:2014,
Bossche:2013,
Kang:2013,
Pelaez:2016,
Ruiz-Alvarez:2015,
Lampe:2012,
Zou:2014,
Thai:2016:DIDC,
Thai:2016:CloudCom,
Sidhanta:2016}, which aims to achieve the desired performance with the minimal cost.
Researchers \cite{Thai:2015:Closer,Oprescu:2012,Thai:2015:Cloud:BoTs} have addressed the problem of \emph{performance maximisation with a budget constraint} with the objective of obtaining with the maximum performance within a budgetary constraint. Chard et al. aimed to help users to \emph{acquire the desired amount of resources with the minimum cost} \cite{Chard:2015}.
Finally, Thai et al.~\cite{Thai:2014:CloudCom} defined a requirement, which focused on making a trade-off between performance and cost. More specifically, the authors asked the user to provide a value, which indicated his or her preference of performance over cost.

\subsection{Parameter Estimation} \label{tax:estimation}

\subsubsection{Overview}

Execution scheduling is the decision-making process that involves different kinds of parameter.
The availability of those factors is crucial since they directly affect the outcome of the scheduling process. For instance, it is impossible to perform cost minimisation scheduling if the cost of cloud resources is unknown.
Some of them are given prior to the execution, e.g. a number of jobs/tasks, while others are unavailable and required to be retrieved or estimated at runtime, i.e. \textbf{parameter estimation}.

\subsubsection{Literature Analysis}

Since a user must pay in order to use cloud resources, the first and foremost parameter that he or she needs to be aware of is the price of cloud resources, i.e. \textbf{monetary factor}.
As discussed earlier, there are three popular pricing schemes for cloud resources: on-demand, spot, and reserved resources. The price of on-demand and reserved resources is always available. However, spot resource pricing is unknown since it changes over time depending on the number of bidders and their bidding prices. As a result, some researchers have decided to set a fixed bidding price, which is lower than the cost of the same amount of on-demand resource.

The \textbf{performance factor} defines the performance of a specific application running on a VM instance type. Notably, the performance factor is specific to each unique pair of application and instance type. For instance, \textbf{task execution time} is the time it takes for a VM to execute one task of an application. By using task execution time, researchers can have a fine-grained control and view of an execution. As a result, this type of parameter is used by the majority of the existing research~\cite{Candeia:2010,
Menache:2014,
Bossche:2013,
Kang:2013,
Duan:2014,
Pelaez:2016,
Oprescu:2010,
Oprescu:2012,
HoseinyFarahabady:2014,
Thai:2015:Cloud:BoT,
Zou:2014,
Thai:2016:DIDC,
Thai:2016:CloudCom,
Thai:2015:Cloud:BoTs}.
Other researchers have used not only the task execution time but also the data transferring time, which denotes the amount of time it takes to transfer data between private and public clouds \cite{Bicer:2012,Thai:2014:CloudCom,Thai:2015:Closer}.
On the other hand, the research of Sidhanta et al. \cite{Sidhanta:2016} represented performance in a more coarse-grained way by using \textbf{job execution time}, i.e., \textbf{makespan}.

The performance factor can also be indirectly represented using different forms. For instance, some existing work used \textbf{resource capacity}, i.e. the number of tasks can be executed by each instance type within a billing cycle, as the performance factor \cite{Ruiz-Alvarez:2015,Mao:2010}.
On the other hand, Chard et al. \cite{Chard:2015} used the queueing or waiting time to indicate that a task could not wait to be executed more than a certain amount of time.

Resource demands can also be used to indirectly represent the performance. More precisely, a resource demand indicates that the desired performance can be achieved if a certain amount of resource is allocated to execute an application.
Resource demand can be represented as the number of VMs required by the application in each billing cycle and has been used by some researchers as a performance factor \cite{Shen:2013,Wang:2016,Lampe:2012}.
On the other hand, other researchers have represented a resource demand as the number of VM hours required to execute an application \cite{Yi:2012,Tang:2012,Yao:2014}.

Most of the time, researchers assume that the performance parameters are available prior to the optimisation process.
However, other researchers have decided not to make this assumption and incorporated the process to estimate the performance parameters as a part of the optimisation process.
For instance, Oprescu et al. \cite{Oprescu:2010}, Thai et al. \cite{Thai:2016:DIDC} and Hoseinyfarahabady et al. \cite{Hoseinyfarahabady:2013} performed a sampling execution, in which a portion of a job was executed on VMs of all available types in order to estimate the task execution time.
On the other hand, Sidhanta et al. \cite{Sidhanta:2016} estimated the job execution time using profiling techniques.

\subsection{Dynamic Scheduling} \label{tax:dynamic}

\subsubsection{Overview}

Section \ref{tax:estimation} has presented the parameters used to schedule a BoT application on the cloud based on a given set of requirements. It is clear that the optimisation decision is made based on the value of those factors. However, these values can dynamically vary during runtime. As a result, the initial scheduling decision may become obsolete and inaccurate. In this section, we discuss the reason for parameter variations and how it is handled by the existing work.

The cost of on-demand and reserved resources remain for the most part constant. However, spot instance prices vary depending on the number of bidders and their bidding prices.

Performance-related parameters can vary during runtime because of one or a combination of the following reasons. As the task's execution time is usually an average value, its actual value can be either greater or less than this average value due to factors such as the input size.
Furthermore, since a cloud is running on top of a heterogeneous cluster consisting of machines of different hardware types, it is possible for VMs of the same type to have different levels of performance since they are located on different hardware infrastructure \cite{O'Loughlin:2014,Ou:2012}.
For instance, Ward and Barker \cite{Ward:2014} showed that the performance of different AWS's VMs of the same type widely varied up to 29\%. Pettijohn et al. \cite{Pettijohn:2014:ICAC} and Chiang et al. \cite{Chiang:2014} also reported such performance variation between VMs on the same type but on different data centres.
Moreover, the performance of the same VM can change over time due to the workload of the physical machine.
Leitner and Cito \cite{Leitner:2014} showed that the performance of IO bandwidth on the same instance could fluctuate up to 30\%.
Moreover, Netflix reported that the performance degradation due to CPU stolen time could be high enough to make it more cost saving to replace a VM by a new one \cite{netflix}.

In order to effectively scale an application during runtime, a user may expect that resources should be available as soon as she requests for them. However, in reality, it normally takes a noticeable amount of time for cloud resources to be made available. This delay referred to as \textbf{instance start-up time}, is common among all cloud providers and can vary from a few seconds up to a few minutes~\cite{Mao:2012:Cloud}. If a user does not take instance startup time into account, the requested resources may be available too late to handle the peak workload.

\subsubsection{Literature Analysis}

As mentioned in the previous section, there are many reasons, which cause parameter variation. This section presents \textbf{dynamic scheduling}, which is performed during runtime in order to handle unexpected events that may result in requirements violation. Our discussion focuses on two aspects of dynamic scheduling: i) how it is triggered and ii) how it is performed.

The simplest way to trigger dynamic scheduling is to perform it periodically, normally right after the monitoring process which updates the parameters to reflect the current state of the execution \cite{Candeia:2010}.
On the other hand, dynamic scheduling can also be triggered periodically at the end of each billing cycle in order to decide if VMs can be terminated for cost saving purpose \cite{Tang:2012,Mao:2010}.

Other researchers have chosen the more specific trigger.
For instance, Bicer et al. \cite{Bicer:2012} proposed to perform dynamic scheduling when each group of jobs was executed.
On the other hand, since Oprescu et al. and Thai et al. \cite{Thai:2015:Closer,Oprescu:2012} aimed to exploit idle VMs, which have no tasks to execute while still being in the current billing cycle, to decrease makespan, the authors started the dynamic scheduling process when an idle VM was detected.
Other researchers have proposed to dynamically reschedule when the requirements are predicted to be violated.
The potentially violated requirements can be constraints such as deadline \cite{Bossche:2013,Kang:2013,Chard:2015,Thai:2016:DIDC,Thai:2016:CloudCom} or budget \cite{Oprescu:2010}.
If spot instances are used, rescheduling is required in case of unexpected termination \cite{Yi:2012,Lu:2013}.
Finally, rescheduling should be performed when the requirements change \cite{Duan:2014}.

Dynamic scheduling can be performed by simply re-running the optimisation process given the updated parameters~\cite{Candeia:2010,Bicer:2012,Yi:2012,Tang:2012,Kang:2013,Duan:2014,Oprescu:2010,Mao:2010}.
However, since it can be time-consuming to perform the whole optimisation process again, there are other, simpler approaches.
For instance, some researchers have focused on re-allocating tasks between VMs in order to improve performance~\cite{Thai:2015:Closer,Oprescu:2012,Thai:2016:DIDC,Thai:2016:CloudCom} while others have proposed approaches to dynamically resize the cluster at runtime~\cite{Lu:2013,Bossche:2013,Chard:2015}.

\subsection{Solving Methods} \label{tax:solving}

\subsubsection{Overview}

Scheduling the execution of BoT job(s) on the cloud is an optimisation problem whose solution, which is typically called \textbf{scheduling plan}, can be found by using the \textbf{solving methods}. Although there are a number of solving methods, they can be grouped into two main categories: the first is \textbf{exact algorithms} which aim to find the \emph{optimal solution} to the problem, i.e. the best scheduling plan possible. The second type of solving methods is \textbf{heuristic algorithms} whose solution may be sub-optimal but can be found quickly.

\subsubsection{Literature Analysis}

The approach of using exact algorithm is proposed in the literature. The scheduling problem is represented as a mathematical model and then solved using an existing solver to find the optimal solution~\cite{Mann:2015}.

The popular approach is using Linear Programming, in which a set of linear formulas are used to model the problem~\cite{Tang:2012}. Integer Linear Programming is another choice which requires all decision variable to be integers~\cite{Candeia:2010,Ruiz-Alvarez:2015,Mao:2010,Thai:2016:DIDC,Thai:2016:CloudCom}, or Binary Integer Programming in which the decision variables are binary values (either one or zero)~\cite{Lampe:2012}.
Other exact algorithms that are reported in the literature include Non-Linear Convex Optimisation Problem~\cite{Sidhanta:2016} and Integer Quadratic Programming~\cite{Thai:2016:CloudCom}.

Exact algorithms guarantee an optimal solution. However, they require a significant amount of time to solve and obtain a scheduling plan. Therefore, these are not suitable for a real-time system in which a decision must be made in a timely manner.

Researchers have adopted the heuristic algorithm approach to reduce the time taken by an exact algorithm. Heuristic algorithms aim to find a solution in a relatively shorter amount of time but do not guarantee global optimality. For generating scheduling plans there may be little difference between local and global optimal solutions.

One of the simplest heuristic approaches is the greedy algorithm which makes the best decision possible given knowledge of the current state.
For instance, scheduling algorithms that iteratively select the cheapest instance type during each iteration have been proposed~\cite{Bossche:2013,Kang:2013,Pelaez:2016,Chard:2015}.
Greedy algorithms which select the best solution given the current states of multiple criteria are proposed~\cite{Thai:2014:CloudCom,Thai:2015:Closer,Wang:2016,Thai:2015:Cloud:BoT,Thai:2015:Cloud:BoTs}.

Heuristic algorithms can incorporate rule-based approaches in which the scheduling decision is based on a set of pre-defined rules. For instance, Gutierrez-Garcia et al.~\cite{Gutierrez-Garcia:2013} use a set of rules defining the order and allocation of tasks to VMs. Other research use rules to acquire resources for achieving the desired performance~\cite{Bicer:2012,Lu:2013,Menache:2014,Shen:2013}.

Another heuristic algorithm is based on dynamic programming, which breaks the scheduling problem into smaller sub-problems~\cite{Yao:2014,Oprescu:2010,Oprescu:2012}.
Meta-heuristic approaches, such as Particle Swarm Optimisation, is a general purpose approach which is employed in this space~\cite{Zou:2014}.

Custom heuristic algorithms are also employed. For instance, Yi et al.~\cite{Yi:2012} use approximation techniques to predict the termination time of spot VMs. The scheduling algorithm of Duan et al. \cite{Duan:2014} is based on the game theory approach. The cloud bursting approach proposed by HoseinyFarahabady et al.~\cite{HoseinyFarahabady:2014} uses the concept of Pareto optimality.

Thai et al.~\cite{Thai:2016:DIDC,Thai:2016:CloudCom} employ a hybrid approach in which an exact algorithm solves different parts of the scheduling problem. Then, the sub-solutions are combined using a heuristic algorithm to produce a complete scheduling plan.

\subsection{Application Heterogeneity} \label{tax:hete}

\subsubsection{Overview}

\textbf{Application heterogeneity} indicates the variety of applications to be scheduled for execution on the cloud.
In other words, methodologies that do not support application heterogeneity are \emph{only able to schedule a single application}.
On the other hand, application heterogeneity is supported when \emph{multiple BoT applications are scheduled at the same time}. Which means that the cloud cluster is shared between multiple applications, each of which performs differently, e.g. has a different task execution time, on the same hardware specification.
Supporting application heterogeneity is challenging since each application prefers a different VM type.
For example, a computation-intensive application prefers a CPU-optimised machine to a memory-optimised instance. As a result, a scheduling mechanism must take into account instance type preferences of all applications in order to select a suitable combination of resource types in a cloud cluster.

\subsubsection{Literature Analysis}

Some researchers have supported application heterogeneity by splitting a cluster into smaller sub-clusters, each of which executes only one application \cite{Menache:2014,Bossche:2013,Chard:2015}. In other words, there is no resource sharing between jobs, i.e. each VM only executes tasks of one job.
However, this approach is inefficient since it limits the flexibility of a cloud cluster. For instance, if only a few tasks of a job are assigned to a VM, it would be wasteful to not use that VM to execute tasks of other jobs.

Resource sharing between applications has been investigated by other researchers.
The simplest approach is to predefine the distribution of jobs on each instance type, as adopted by Mao et al. \cite{Mao:2010}. Alternative approaches have been proposed, which assign a group of jobs, instead of just a single one, to be executed on a sub-cluster \cite{Yao:2014,HoseinyFarahabady:2014}. The authors have developed mechanisms which create a group of jobs so that the resource wastefulness in each sub-cluster can be minimised.
Finally, the most complicated but also most efficient approach is to assign all jobs to all instances without predefined task distribution or job grouping~\cite{Shen:2013,
Wang:2016,
Kang:2013,
Duan:2014,
Pelaez:2016,
Thai:2015:Cloud:BoT,
Lampe:2012,
Gutierrez-Garcia:2013,
Zou:2014,
Thai:2016:DIDC,
Thai:2016:CloudCom,
Thai:2015:Cloud:BoTs}. This approach results in a workload assignment in which each VM receives a different task distribution. As a result, it can potentially maximise resource efficiency. However, this approach is challenging since it may have to consider countless possibilities of workload allocation between jobs and VMs.

\section{Discussion}
\label{sec:discussion}
Based on the survey of existing work presented in Section~\ref{sec:existing} and the taxonomy developed in Section~\ref{sec:taxonomy}, this section will summarise the current trends in this research area. Further, we propose research directions that will improve the usage of cloud resources.

\subsection{Current Trends}
\label{discussion:currenttrends}

\afterpage{
\begin{landscape}
    \begin{table}
    \centering
    \caption{Taxonomy of BoT Scheduling Methodologies}
    \label{tbl:taxo}
    \begin{tabular}{|l|c|c|c|c|c|c|c|c|c|c|c|c|}
    \hline
    \multirow{3}{*}{} & \multicolumn{3}{c|}{\multirow{2}{*}{Functionality}} & \multicolumn{4}{c|}{Requirement} & \multirow{3}{*}{\begin{tabular}[c]{@{}c@{}}Dynamic\\ Scheduling\end{tabular}} & \multirow{3}{*}{\begin{tabular}[c]{@{}c@{}}Parameter\\ Estimation\end{tabular}} & \multicolumn{2}{c|}{\multirow{2}{*}{\begin{tabular}[c]{@{}c@{}}Solving\\ Method\end{tabular}}} & \multirow{3}{*}{\begin{tabular}[c]{@{}c@{}}Application\\ Heterogeneity\end{tabular}} \\ \cline{5-8}
     & \multicolumn{3}{c|}{} & \multicolumn{2}{c|}{Constraint} & \multicolumn{2}{c|}{Objective} &   &   & \multicolumn{2}{c|}{} &   \\ \cline{2-8} \cline{11-12}
     & \begin{tabular}[c]{@{}c@{}}Type\\ Selection\end{tabular} & \begin{tabular}[c]{@{}c@{}}Resource\\ Scaling\end{tabular} & \begin{tabular}[c]{@{}c@{}}Workload\\ Allocation\end{tabular} & Cost & Perf & Cost & Perf &   &   & Exact & Heur &   \\ \hline
    \cite{Bicer:2012}                         &   & X &   & X & X & X & X &   &   &   & X &   \\ \hline
    \cite{Yi:2012}                            &   & X &   &   &   & X & X &   & X & X &   &   \\ \hline
    \cite{Lu:2013}                            &   & X &   & X & X &   &   & X &   & X &   &   \\ \hline
    \cite{Duan:2014:CloudCom}                          &   & X & X & X & X &   &   &   &   & X &   & X \\ \hline
    \cite{Tang:2012}                          &   & X & X &   & X & X &   & X &   &   & X &   \\ \hline
    \cite{Candeia:2010}                       &   & X & X &   & X & X &   & X &   & X &   &   \\ \hline
    \cite{Menache:2014}                       &   & X & X &   & X & X &   & X &   &   &   & X \\ \hline
    \cite{Yao:2014}                           &   & X & X &   & X & X &   &   &   &   & X & X \\ \hline
    \cite{Shen:2013}                          &   & X & X &   & X & X &   &   &   &   & X & X \\ \hline
    \cite{Mao:2010}                           & X & X &   & X & X &   &   &   &   & X &   & X \\ \hline
    \cite{Thai:2014:CloudCom}                 & X & X & X & X & X &   &   &   &   &   & X &   \\ \hline
    \cite{HoseinyFarahabady:2014}             & X & X & X & X & X &   &   &   &   &   & X & X \\ \hline
    \cite{Kang:2013}                          & X & X & X & X & X & X & X & X &   &   & X & X \\ \hline
    \cite{Ruiz-Alvarez:2015}                  & X & X &   &   &   & X & X &   &   & X &   &   \\ \hline
    \cite{Duan:2014}                          & X & X & X &   &   & X & X &   &   &   & X & X \\ \hline
    \cite{Gutierrez-Garcia:2013}              & X & X & X &   &   & X & X &   &   &   & X & X \\ \hline
    \cite{Thai:2015:Closer}                   & X & X & X & X &   &   & X & X &   &   & X &   \\ \hline
    \cite{Thai:2015:Cloud:BoTs}               & X & X & X & X &   &   & X &   &   &   & X & X \\ \hline
    \cite{Oprescu:2010,Oprescu:2012}          & X & X & X & X &   &   & X & X & X &   & X &   \\ \hline
    \cite{Bossche:2013}                       & X & X & X &   & X & X &   & X &   &   & X & X \\ \hline
    \cite{Lampe:2012}                         & X & X & X &   & X & X &   &   &   & X & X & X \\ \hline
    \cite{Zou:2014}                           & X & X & X &   & X & X &   &   &   &   & X & X \\ \hline
    \cite{Sidhanta:2016}                      & X & X &   &   & X & X &   &   & X & X &   &   \\ \hline
    \cite{Chard:2015}                         & X & X & X &   & X & X &   & X &   &   & X & X \\ \hline
    \cite{Wang:2016}                          & X & X & X &   & X & X &   &   &   &   & X & X \\ \hline
    \cite{Pelaez:2016}                        & X & X & X &   & X & X &   & X &   &   & X & X \\ \hline
    \cite{Thai:2015:Cloud:BoT}                & X & X & X &   & X & X &   &   &   &   & X &   \\ \hline
    \cite{Thai:2016:DIDC,Thai:2016:CloudCom}  & X & X & X &   & X & X &   & X & X & X & X & X \\ \hline
    \cite{Hoseinyfarahabady:2013}             & X & X & X & X & X &   &   & X & X & X &   &   \\ \hline
    \end{tabular}
    \end{table}
\end{landscape}
}

Table~\ref{tbl:taxo} summarises all methods reviewed in this paper and categorises them using our proposed taxonomy.

\textbf{Functionality}: Resource scaling is a key feature that is supported by all research. It is noted that type selection can only be obtained on frameworks that support heterogeneous cloud environments.
The majority of existing research supports workload allocation since it provides fine-grained control over task distribution between jobs and VMs.

\begin{itemize}
\item \textbf{Requirements}: performance constraints (for example, deadline constraint) with cost objectives (for example, cost minimisation) is the most popular requirement. This is because a desirable level of performance needs to be achieved while being aware of the monetary costs incurred in real time. This is unique to using cloud resources for executing an application in contrast to grids or in-house clusters.

\item \textbf{Dynamic Scheduling}: less than half of the surveyed research publications support dynamic scheduling. For the sake of simplicity, a number of publications assume that the performance of cloud computing resources remains unchanged during execution. However, this assumption does not hold for real clouds and when heterogeneous resources are used. 

\item \textbf{Parameter Estimation}: 
only four papers we surveyed offer any support for parameter estimation. This reflects the common belief that the necessary parameters can be obtained prior to executing a job. However, obtaining this information prior to execution is not always feasible given that the environment is usually shared between a number of users.

\item \textbf{Solving Method}: the majority of research adopts a heuristic approach since it provides a solution faster than alternative approaches, such as exact algorithms. Approaches that are required for real-time systems need to converge on a scheduling plan as quickly as possible. Although exact algorithms are guaranteed to find an optimal solution, they are not widely used since they are time-consuming. We found that exact algorithms are adopted in only seven research papers we surveyed.

\item \textbf{Application Heterogeneity}: the vast majority of existing research supports application heterogeneity - scheduling the execution of multiple applications that perform differently on the same type of VM. This reflects the characteristic of cloud environments that can be shared between different users with a wide range of applications.

\end{itemize}

Based on Table~\ref{tbl:taxo}, we summarise the current trends in optimising the use of cloud resources as follows:
\begin{itemize}
    \item \textbf{Supporting heterogeneous cloud environments:} cloud computing environments need to be flexible in accommodating multiple applications with diverse needs by supporting VMs with different hardware specification. Therefore, cloud environments are supporting VMs of different instance types.
    \item \textbf{Minimising monetary cost while ensuring the desired performance:} monetary costs is one of the most important concerns for cloud users. Hence, there is research to produce scheduling plans that keep the cost as low as possible without sacrificing the desired quality of service.
    \item \textbf{Handling unexpected events at runtime:} in any large scale and real-time system, unexpected events, such as missing information or performance variation during runtime, occur inevitably. Hence, mechanisms are put in place to detect and handle such events in order to prevent, or at least minimise their impact.
    \item \textbf{Using heuristic algorithms:} heuristic algorithms are popularly used for optimising resources for the execution of BoT applications. This is because they can produce timely results although there is a trade-off against the optimality of solutions obtained.
\end{itemize}

\subsection{Future Directions}
\label{discussion:futuredirections}

We present the following three avenues as directions for future research to bring the area of optimising cloud usage for executing BoT applications to maturity and to develop next-generation cloud computing systems.

\begin{itemize}
\item \textbf{Heterogeneous Physical Architecture Awareness:} performance variation can be caused by the heterogeneity of the underlying hardware in the cloud. Currently, one practical solution that is used to mitigate this at the middleware level, includes dynamic scheduling for reassigning tasks between VMs. The next step will be to incorporate flexible mechanisms in scheduling that selects resources instead of different instance types by determining the underlying hardware suitable for executing an application. For instance, Ou et al.~\cite{Ou:2012} propose an approach which achieves performance improvements without resulting in any additional cost by probabilistically selecting VMs with better performance.

\item \textbf{Cross Cloud Scheduling:} mature research in optimisation of cloud usage focuses on using only one cloud provider. In order to avoid vendor lock-in, there are mechanisms that allow for deployment of applications across multiple providers (for example~\cite{crosscloudbot-1}). This creates a more flexible environment in which users can actively select the most cost-effective cloud to run their application. However, increasing the number of cloud providers consequently increases the number of options that the scheduling process must consider, which significantly expands the search space. This also requires scheduling frameworks that are not provider-specific, but are compatible with different standards of multiple providers.

\item \textbf{Performance Prediction:} parameter estimation should receive more attention from the research community. There should be more focus on performance prediction rather than simply sampling applications. This will reduce the overheads in creating new VMs and improving the accuracy of estimation. Predicting the performance of an application can be challenging since it requires sophisticated profiling techniques to analyse an application statically or dynamically.

\item \textbf{Optimising the Usage of Reserved Instances:} it is obvious that there is a lack of existing research in this area on reserved instances. More specifically, only 2 out of the 31 selected publications focused on the pricing scheme corresponding to reserved instances. This is perhaps due to the commitment (for example, for 1 to 3 years) of reserving resources upfront. However, we believe that this avenue should not be ignored, given that many middle and large organisations are moving their operations to the cloud (for instance, most of the Netflix streaming service is hosted via the AWS cloud and it would not be difficult for such a service to commit to the utilisation of cloud resources).

\end{itemize}

\section{Conclusions}
\label{sec:conclusions}

Cloud computing provides a flexible environment to execute BoT applications, by offering on-demand resources, which can be seamlessly provisioned at runtime. However, to use cloud computing resources effectively requires scheduling approaches that consider not only the requirements of the users but also the performance of the applications.

In this survey, we have reviewed the existing publications in optimising cloud usage for executing BoT applications. From this review, we constructed a taxonomy describing different aspects and characteristics of this research area, such that frameworks can be compared against one another. By applying the proposed taxonomy to existing research, we inferred that the current research trend is to build heterogeneous and flexible cloud clusters that (i) achieve the desired quality of service with minimum costs and (ii) handle unexpected events occurring during execution. To conclude we suggest future research directions, in order to improve the quality of cloud usage and addresses the challenges arising in real-world scenarios.

This survey is far from providing a complete view of cloud usage optimisation. For future work, we are planning to survey existing research for other types of application besides BoT, such as workflow \cite{Barker:2007}, MapReduce \cite{Dean:2008} and user-facing \cite{Cirne:2013,Nikolay:2014} applications. 

\bibliographystyle{elsarticle-num}
\bibliography{references}

\end{document}